\documentstyle[aps,epsf,amssymb]{revtex}
\begin{document}
\twocolumn
\title{{\scriptsize Phys. Rev. A {\bf 65}, 062319
(2002).}
\\Limits to clock synchronization induced by \\ completely
dephasing communication channels}
\author{V. Giovannetti,$^*$ S. Lloyd,$^{1,\dag}$
L. Maccone,$^\ddagger$ and M. S. Shahriar$^\S$}
\address{Massachusetts Institute of Technology, Research Laboratory
of Electronics, 50 Vassar St.\\ $^1$Department of Mechanical
Engineering,\\ Cambridge, MA 02139, USA.}
\maketitle
\begin{abstract}
{\bf Abstract.}  Clock synchronization procedures are analyzed in the
presence of imperfect communications. In this context we show that
there are physical limitations which prevent one from synchronizing
distant clocks when the intervening medium is completely dephasing, as 
in the case of a rapidly varying dispersive medium.
\end{abstract}
\bigskip
There are two main kinds of protocols for achieving clock
synchronization. The first is ``Einstein synchronization protocol''
{\cite{einstsyn}} in which a signal is sent back and forth between one
of the clocks (say Alice's clock) and the other clocks. By knowing the
signal speed dependence on the intermediate environment, it is
possible to synchronize all the clocks with Alice's. The other main
protocol is ``Eddington slow clock transfer'' {\cite{eddinsyn}}: after
locally synchronizing it with hers, Alice sends a clock ({\it i.e.} a
physical system that evolves in time with a known time dependence) to
all the other parties. The clocks transfer must be of course perfectly
controllable, as one must be able to predict how the clock will react
to the physical conditions encountered {\it en route} which may shift
its time evolution. Moreover, since any acceleration of the transferred
clocks introduces a delay because of relativistic effects, one must
suppose that the transfer is performed `adiabatically slowly', {\it
i.e.}  such that all accelerations are negligible. Notice that the
above protocols can be implemented using only classical resources:
peculiar quantum features such as entanglement, squeezing, {\it etc},
are not needed. In what follows, such synchronization schemes will be
referred to as `classical protocols'.

A recently proposed quantum clock synchronization protocol
{\cite{jozsa}} was found {\cite{comment}} to be equivalent to the
Eddington slow clock synchronization.  The application of entanglement
purification to improve quantum clock synchronization in presence of
dephasing was attempted without success in {\cite{preskill}}. One
might think there are other ways to implement a synchronization scheme
that employ quantum features such as entanglement and squeezing, but
this paper shows that this is not the case. In fact, it will be shown
that quantum mechanics does not allow one to synchronize clocks if it
would not be possible to employ also one of the classical protocols,
which one can always employ if the channel is perfect or if its
characteristics are controllable. However, the relevance of quantum
mechanics in the clock synchronization procedures should not be
underestimated, since there exist schemes that exploit quantum
mechanics to achieve a (non-classically allowed) increase in the
accuracy of classical clock synchronization protocols, such as the one
obtainable exploiting entangled systems {\cite{chuang,altro,prl}}.

The presented discussion takes also into account the possibility that
the two distant parties who want to synchronize their clocks (say
Alice and Bob) and who are localized in space, can entangle their
systems by exchanging a certain number of quantum states and the
possibility that they may employ the ``wave function collapse''
{\cite{jozsa}}, through post-selection measurements. The intuitive
idea behind the proof is as follows. To synchronize clocks, Alice and
Bob must exchange physical systems such as clocks or pulses of light
that include timing information. But any effect, such as rapidly
varying dispersion, that randomizes the relative phases between energy
eigenstates of such systems completely destroys the timing
information. Any residual information, such as entanglement between
states with the same energy, cannot be used to synchronize clocks as
shown below.\par

The paper is organized as follows. In Sect. {\ref{teorema}} the
analytic framework is established. In Sect. {\ref{cs}} the clock
synchronization procedure is defined and the main result is
derived. In particular, in Subsect. {\ref{dimostraz}} the exchange of
quantum information between Alice and Bob is analyzed and in
Subsect. {\ref{ps}} the analysis is extended to include partial
measurements and post-selection schemes in the synchronization
process.

\section{The system}\label{teorema}
Assume the following hypotheses that describe the most general
situation in which two distant parties communicate through a noisy
environment:\begin{enumerate}\item\label{local} Alice and Bob are {\it
separate} entities that initially are {\it disjoint}. They belong to
the same inertial reference frame and communicate by exchanging some
physical system.\item\label{dephasing} The environment randomizes the
phases between different energy eigenstates of the exchanged system
while in transit.\end{enumerate} From these hypotheses it will be
shown that Alice and Bob cannot synchronize their clocks.\par

In Subsect. {\ref{s:ip1}} we explain the first hypothesis by
giving its formal consequences. In Subsect. {\ref{s:ip2}} we analyze
the second hypothesis and explain how it describes a dephasing
channel. 

\subsection{First hypothesis}\label{s:ip1}
The first hypothesis states the problem and ensures that {initially}
Alice and Bob do not already share any kind of system that acts as a
synchronized clock. By {\it separate} we mean that at any given time
Alice and Bob cannot gain access to the same degrees of freedom and
there is no direct interaction between Alice and Bob's systems. This
can be described by the following properties of the system's Hilbert
space and Hamiltonian. At time $t$ the Hilbert space of the global
system can be written as
\begin{eqnarray} 
{\cal H}={\cal H}_A(t)\otimes{\cal H}_C(t)\otimes{\cal H}_B(t)
\;\label{spazio},
\end{eqnarray}
where the Hilbert space ${\cal H}_A(t)$ refers to the system on which
Alice can operate at time $t$, ${\cal H}_B(t)$ refers to Bob's system,
and ${\cal H}_C(t)$ describes the systems on which neither of them can
operate. The time dependence in Eq. (\ref{spazio}) does not imply that
the global Hilbert space changes in time, but it refers to the
possibility that a system that was previously under Alice's influence
has been transferred to Bob (or {\it viceversa}), after a transient
time in which it cannot be accessed by any of them. Since information
{\it must} be encoded into a physical system, this mechanism describes
any possible communication between them.  Moreover, the Hamiltonian of
the system can be written as
\begin{eqnarray} H(t)=H_A(t)+H_B(t)+H_C(t)
\;\label{hamilttot},
\end{eqnarray}
where the time dependent $H_A(t)$ and $H_B(t)$ evolve the states in
${\cal H}_A$ and ${\cal H}_B$ under the control of Alice and Bob
respectively, while $H_C(t)$ evolves the system in transit between
them when it is not accessible.  As a consequence of
Eq. (\ref{spazio}), at time $t$ the three terms on the right side of
Eq. (\ref{hamilttot}) commute, since they act on different Hilbert
spaces. For the same reason any operator under the influence of Alice
at time $t$ commutes with all Bob's operator at the same time.  A
simple example may help explain this formalism.  Consider the
situation in which the system is composed of three 1/2 spin particles
(qubits). A possible communication is then modeled by the sequence
\begin{figure}[hbt]
\begin{center}\epsfxsize=.8
\hsize\leavevmode\epsffile{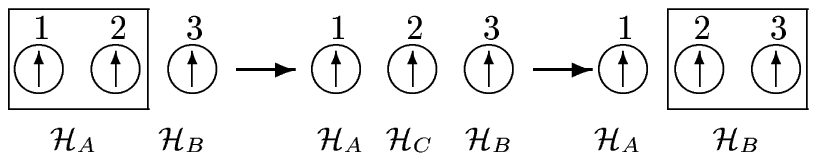} 
\end{center}
\vspace{-.5cm}
\end{figure}
{\it i.e.} initially Alice's Hilbert space ${\cal H}_A$ contains spins
1 and 2, and Bob owns only spin 3. Alice then encodes some information
on spin 2 (eventually entangling it with spin 1), and sends it to
Bob. There will be a time interval in which none of them can access
spin 2 and this situation corresponds to having spin 2 belonging to
${\cal H}_C$. Finally, Bob receives spin 2, and his Hilbert space
${\cal H}_B$ describes both spins 2 and 3. Notice that the form of the
Hamiltonian in Eq. (\ref{hamilttot}), where no interaction terms are
present, allows each of them to act, at a given time $t$, only on the
spins that live in their own Hilbert space at time $t$.  An analogous
description applies also to more complicated scenarios, such as the
exchange of light pulses. In this case, causality constraints allow
Alice and Bob to act only on localized traveling wave modes of the
electromagnetic field. Thus, also here, it is possible to define a
traveling system Hilbert space ${\cal H}_C$ that factorizes as in
Eq. (\ref{spazio}). From the above example, it is easy to see that in
each communication exchange, it is possible to define a {\it
departure} time $t_s$ after which the sender cannot act any more on
the system in transit, and an {\it arrival} time $t_r$ before which
the receiver cannot yet act on such system. It is between these two
times that the exchanged system belongs to ${\cal H}_C$.

In Hyp.~{\ref{local}} by initially {\it disjoint} we mean that Alice
and Bob do not share any information prior to the first communication
exchange. In particular this means that, before they start to
interact, the state of the system factorizes as
\begin{eqnarray}
|\Psi\rangle=|\phi\rangle_A\otimes|\varphi\rangle_B
\;\label{statoiniz},
\end{eqnarray}
{\it i.e.} the initial state is not entangled and they do not share
any quantum information. Here $|\phi\rangle_A$ is the state of Alice's
system evaluated at the time at which she starts to act, while
$|\varphi\rangle_B$ is the state of Bob's system evaluated at the time
at which {\it he} starts to act. For ease of notation, the tensor
product symbol $\otimes$ will be omitted in the following except when
its explicit presence helps the comprehension.

\subsection{Second hypothesis}\label{s:ip2}
The second hypothesis imposes limitations to the information retrieved
from the exchanged signal.  The dephasing of the energy eigenstates
describes the non-dissipative noise present in most non-ideal
communication channels and implies a certain degree of decoherence in
any quantum communication between Alice and Bob. Define $|e,d\rangle$
as the eigenstate relative to the eigenvalue $\hbar\>\omega_e$ of the
free Hamiltonian of the exchanged system $C$. The label $d$ takes into
account possible degeneracy of such eigenstate. We assume that during
the travel, when neither Alice nor Bob can control the exchanged
system in ${\cal H}_C$, the states $|e,d\rangle$ undergo the
transformation
\begin{eqnarray}
|e,d\rangle\longrightarrow e^{-i\varphi_e}|e,d\rangle
\;\label{evolener},
\end{eqnarray}
where the random phase $\varphi_e\in[0,2\pi[$ is independent of $d$.
The channel dephasing arises when different energy eigenstates are
affected by different phase factors $\varphi_e$. For this reason the
dephasing is characterized by the joint probability function
$p_\epsilon(\varphi_e,\varphi_{e'})$, that weights the probability
that the energy levels $|e,d\rangle$ and $|e',d\rangle$ are affected
by the phases $\varphi_e$ and $\varphi_{e'}$ respectively. The
parameter $\epsilon\in[0,1]$ measures the degree of decoherence in the
channel. In particular, $\epsilon=1$ describes the case of complete
decoherence, where the phases relative to different energy eigenstates
are completely uncorrelated, namely
$p_\epsilon(\varphi_e,\varphi_{e'})$ is a constant. On the other hand,
$\epsilon=0$ describes the case of no decoherence, where each energy
eigenstate acquires the same phase, namely
$p_\epsilon(\varphi_e,\varphi_{e'})\to\delta(\varphi_e-\varphi_{e'})/2\pi$.
Written in the energy representation, the channel density matrix
$\varrho_c$ evolves, using Eq.~(\ref{evolener}), as 
\begin{eqnarray}
\varrho_c=\sum_{e\;e'}P_e\varrho_cP_{e'}\quad\longrightarrow\quad
\sum_{e\;e'}e^{-i(\varphi_e-\varphi_{e'})}\;P_e\varrho_cP_{e'}
\;\label{ev},
\end{eqnarray}
where $P_e=\sum_d |e,d\rangle\langle e,d|$ is the projection operator
on the channel eigenspace of energy $\hbar\>\omega_e$. Taking into
account the stocasticity of the evolution (\ref{evolener}), the right
hand term of Eq.~(\ref{ev}) must be weighted by the probability
distribution $p_\epsilon(\varphi_e,\varphi_{e'})$, resulting in
\begin{eqnarray}
\varrho_c\longrightarrow\sum_{e\;e'}\delta^{(\epsilon)}_{ee'}
P_e\;\varrho_c\;P_{e'} 
\;\label{evolchan},
\end{eqnarray}
where \begin{eqnarray}
\delta^{(\epsilon)}_{ee'}=\int_0^{2\pi}d\varphi_e\int_0^{2\pi}
d\varphi_{e'} \;
p_\epsilon(\varphi_e,\varphi_{e'})\;e^{-i(\varphi_e-\varphi_{e'})}
\;\label{deltadef}.
\end{eqnarray}  The width of the function
$\delta^{(\epsilon)}_{ee'}$ decreases with $\epsilon$, so that
$\delta^{(\epsilon=0)}_{ee'}$ is independent of $e$ and $e'$ and the
state is unchanged, while $\delta^{(\epsilon=1)}_{ee'}$ is the
Kronecker delta and the state suffers from decoherence in the energy
eigenstate basis.

The dephasing process of Eq. (\ref{evolener}) can be derived assuming
a time dependent Hamiltonian $H_C(t)=H^o_C+H'_C(t)$, where $H^o_C$ is
the free evolution of the system with eigenstates $|e,d\rangle$ and
$H'_C(t)$ is a stochastic contribution that acts on the system in a
small time interval $\delta t$ by shifting its energy eigenvalues by a
random amount $\nu_e$, such that $\nu_e\delta t=\varphi_e$. In fact,
in the limit $\delta t\to 0$, the evolution of the exchanged system is
described by
\begin{eqnarray} 
U_C(t_{r},t_{s})=\exp[{-i\sum_{e}P_e
\left(\omega_e(t_{r}-t_{s})+\varphi_e\right)}
]
\;\label{udic}, 
\end{eqnarray}
where $\hbar\>\omega_e$ is the energy eigenvalue of the exchanged
system relative to the eigenvector $|e,d\rangle$ and $t_{r}$ and
$t_{s}$ are the exchanged system's arrival and departure times
respectively introduced in Subsect. {\ref{s:ip1}}
{\cite{nota}}. Notice that for $\varphi_e$ independent on $e$ (which
corresponds to the case $\epsilon=0$), $U_C$ reduces to the
deterministic free evolution operator $\exp[-\frac i\hbar
H^o_C(t_r-t_s)]$, apart from an overall phase term.

It might be interesting to consider the simpler case in which the
random phase $\varphi_e$ can be written as $\omega_e\theta$ with the
random term $\theta$ independent on $e$. In this case, Eq. (\ref{udic})
simplifies to \begin{eqnarray} U_C(t_{r},t_{s})=\exp[
{-\frac i\hbar
H^o_C(t_r-t_s+\theta)}]
\;\label{udic2}.
\end{eqnarray}
This last situation depicts the case in which all signals exchanged
between Alice and Bob are delayed by an amount $\theta$. As an example
consider light signals which encounter a medium with unknown (possibly
varying) refractive index or a traveling `clock' which acquires an
unpredictable delay. The situation described by (\ref{udic}) is even
worse, since not only such a delay may be present, but also the wave
function of the system is degraded by dispersion effects. In both
cases, the information on the transit time $t_r-t_s$ that may be
extracted from $U_C(t_r,t_s)$ depends on the degree of randomness of
$\varphi_e$. In particular, if $\varphi_e$ is a completely random
quantity ({\it i.e.}  for $\epsilon=1$), no information on the transit
time can be obtained. 

This, of course, prevents the possibility of using classical
synchronization protocols, where unknown delays in either the signal
travel time or in the exchanged clock prove to be fatal. One might
think that exploiting the apparently non-local properties of quantum
mechanics ({\it e.g.} entanglement), these limits can be overcome. In
the following sections we will show that this is not the case.

\section{Clock synchronization}\label{cs}
In this section we analyze the clock synchronization schemes in detail
and show the effect of a dephasing communication channel. 

\begin{figure}[hbt]
\begin{center}\epsfxsize=.8
\hsize\leavevmode\epsffile{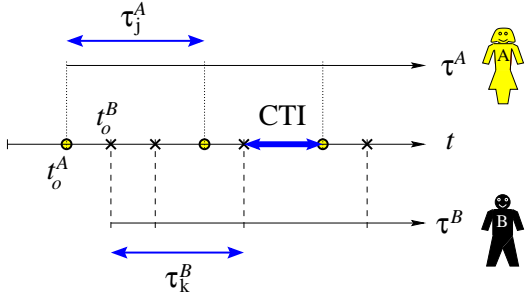} 
\end{center}
\caption{Comparison between the times $\tau^A$ and $\tau^B$ of Alice
and Bob's clocks. The center line represents the ``absolute'' time as
measured by an external clock. The small circles represent the times
of events that take place on Alice's side, while the crosses represent
those on Bob's side. The upper line is the time as measured by Alice's
clock: she only has direct access to the proper time intervals such as
$\tau_j^A$. Analogously, the lower line represents Bob's proper
time. To achieve clock synchronization, Alice and Bob need to recover
a connecting time interval (CTI) such as the one shown. }
\label{f:events}\end{figure}

How does synchronization take place? Define $t_0^A$ and $t_0^B$ as the
initial times of Alice and Bob's clocks as measured by an external
clock. (Of course, since they do not have a synchronized clock to
start with, they cannot measure $t_0^A$ and $t_0^B$.) Alice and Bob
will be able to synchronize their clocks {\it if and only if} they can
recover the quantity $t_0^A-t_0^B$, or any other time interval that
connects two events that happen one on Alice's side and the other on
Bob's side. Each of them has access to the times at which events on
her/his side happen and can measure such events only relative to their
own clocks. We will refer to these quantities as `proper time
intervals' (PTIs). For Alice such quantities are defined as
$\tau_j^A=t_j^A-t_0^A$, where $t_j^A$ is the time at which the $j$-th
event took place as measured by the external clock. Analogously for
Bob we define his PTI as $\tau_k^B=t_k^B-t_0^B$. If Alice and Bob
share the data regarding their own PTIs, they cannot achieve
synchronization: they need also a `connecting time interval' (CTI),
{\it i.e.} a time interval that connects an event that took place on
Alice side with an event that took place on Bob's side as shown in
Fig. {\ref{f:events}}.

In this framework, consider the case of Einstein and Eddington's clock
synchronization. In Einstein clock synchronization the PTIs on Alice
side are the two times at which she sent and received back the signal
she sends Bob. Bob's PTI is the time at which he bounces back the
signal to Alice. The CTI in this case measures the time difference
between the events ``Alice sends the signal'' and ``Bob bounces the
signal back''. The protocol allows Alice to recover the CTI by simply
dividing by two the time difference between her two PTIs. The analysis
of Eddington's slow clock transfer is even simpler. In this case Bob's
PTI is the time at which Bob looks at the clock Alice has sent him
after synchronizing it with hers. The CTI is, for example, the time
difference between the event ``Bob looks at the clock sent by Alice''
and ``on Alice side it's noon'': Bob can recover it just looking at
the time shown on the clock he received from Alice.

In this paper we show that in the presence of a dephasing
communication channel (as described in Hyp.~{\ref{dephasing}}), there
is no way in which Alice and Bob may achieve a CTI. The best that they
can do is to collect a series of PTIs related to different events and
a collection of CTI transit times corrupted by the noisy communication
line: clock synchronization is thus impossible.

\subsection{Timing information exchange}\label{dimostraz}
In this section we analyze the exchange of quantum information between
Alice and Bob in the presence of dephasing.

Starting from the state $|\Psi\rangle$ of Eq. (\ref{statoiniz}), Alice
and Bob begin to act on their systems at two times (that are not
necessarily the same), in order to get ready for the information
transfer.  Without loss of generality one can assume that these two
times coincide with their own time origins, {\it i.e.}  $t_0^A$ and
$t_0^B$. This means that, at those two times, they introduce time
dependent terms in the system Hamiltonian:
\begin{eqnarray} &&H^o_A\ \longrightarrow\ H_A(t)\equiv
H^o_A+H_A'(t-t_0^A)\;\nonumber\\ &&H^o_B\ \longrightarrow\
H_B(t)\equiv H^o_B+H_B'(t-t_0^B)\;\label{hamilta},
\end{eqnarray}
where $H^o_A$ and $H^o_B$ are the free Hamiltonians of Alice and Bob's
systems and $H_A'(t-t_0^A)$ and $H_B'(t-t_0^B)$ characterize the most
general unitary transformations that they can apply to their
systems. These last terms are null for $t<t_0^A$ and $t<t_0^B$ (when
they haven't yet started to act on their systems). Notice that
according to Eq. (\ref{spazio}), also the domains of $H_A(t)$ and
$H_B(t)$ may depend on time.

Suppose first that Alice is going to send a signal to Bob. Define
$t_s^A$ the departure time at which Alice sends a message to Bob
encoding it on a system described by the Hilbert space ${\cal
H}_c$. This implies that the system she has access to will be ${\cal
H}_a$ up to $t_s^A$ and ${\cal H}_{a'}$ afterward, so that ${\cal
H}_a={\cal H}_{a'}\otimes{\cal H}_c$. In the same way, defining
$t_r^B$ as the arrival time on Bob's side, we may introduce a space
${\cal H}_{b'}={\cal H}_{b}\otimes{\cal H}_c$ that describes the
Hilbert space on which Bob acts after $t_r^B$. The label $A$ on
$t_s^A$ refers to the fact that the event of sending the message
happens locally on Alice's side, so in principle she can measure such
a quantity as referred to her clock as the PTI
$\tau_s^A=t_s^A-t_0^A$. Analogous consideration applies to Bob's
receiving time $t_r^B$ and Bob's PTI $\tau^B_r=t_r^B-t_0^B$.

\begin{figure}[hbt]
\begin{center}\epsfxsize=.8
\hsize\leavevmode\epsffile{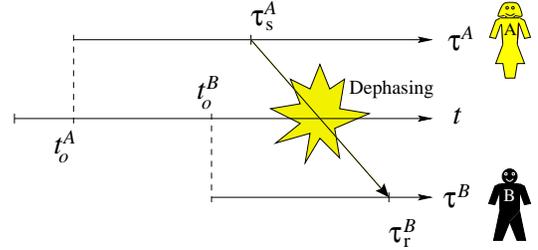} 
\end{center}
\caption{Alice sends Bob a message encoded into a quantum system $C$
at time $t_s^A$ (her proper time $\tau_s^A$) and Bob receives it at
time $t_r^B$ (his proper time $\tau_r^B$). During the travel the
system $C$ undergoes to dephasing. }
\label{f:times}\end{figure}
Consider the situation of Fig. {\ref{f:times}} in which, for
explanatory purposes, $t_0^A<t_0^B<t_s^A<t_r^B$. Start from the group
property of the time evolution operators
\begin{eqnarray}
U(t,0)=U(t,t')\;U(t',0)\;\label{groupprop},
\end{eqnarray}
and the commutativity of the operators that act on the distinct spaces
of Alice and Bob. It's easy to show that for $t_s^A\leqslant
t\leqslant t_r^B$ the state of the system is given by \begin{eqnarray}
|\Psi(t)\rangle&=&U_b(t,t_0^B)\times\label{evol1}\\\nonumber&&
U_{a'}(t,t_s^A)\;U_c(t,t_s^A)\;
U_a(t_s^A,t_0^A)\;|\Psi\rangle
\;,
\end{eqnarray}
where $U_x(t,t')$ is the evolution operator in space ${\cal H}_x$ and
\begin{eqnarray} |\Psi\rangle\equiv U_a(t_0^A,0)\;U_b(t_0^B,0)\;
|\Psi(0)\rangle
\;\label{statoiniziale}
\end{eqnarray}
is the initial state as far as Alice and Bob are concerned, defined in
Eq.  (\ref{statoiniz}). By Hyp.~{\ref{local}} this state does not
contain any usable information on $t_0^A$ and $t_0^B$. In
Eq. (\ref{evol1}) notice that up to time $t_s^A$ the systems ${\cal
H}_c$ and ${\cal H}_{a'}$ are evolved together by $U_a$. Analogously,
for $t\geqslant t_r^B$ after Bob has received the system Alice sent
him, one has
\begin{eqnarray}
|\Psi(t)\rangle=U_{a'}(t,t_r^B)\;U_{b'}(t,t_r^B)\;|\Psi(t_r^B)\rangle
\;\label{evol2}.
\end{eqnarray}
Joining (\ref{evol1}) and (\ref{evol2}), it follows
\begin{eqnarray}
|\Psi(t)\rangle&=&U_{b'}(t,t_r^B)\;U_b(t_r^B,t_0^B)
\;\times\nonumber\\
&&U_{a'}(t,t_s^A)\; U_c(t_r^B,t_s^A)\; U_a(t_s^A,t_0^A)\;|\Psi\rangle
\;\label{evol3}.
\end{eqnarray}

The time dependence of Alice and Bob's Hamiltonians (\ref{hamilta})
allows to write their unitary evolution operators as functions of
their PTIs, {\it i.e.}\begin{eqnarray}
&&U_\alpha(t',t'')=\;\stackrel{\longleftarrow}
{\exp}\left[-\frac i\hbar{\int_{t''}^{t'}dt\;[H^o_\alpha+
H'_\alpha(t-t_0^A)]}\right]=
\label{ubarra}\\\nonumber&&\stackrel{\longleftarrow}
{\exp}\left[-\frac i\hbar{\int_{t''-t_0^A}^{t'-t_0^A}dt\;
[H^o_\alpha+H'_\alpha(t)]}\right]\equiv \bar
U_\alpha(\tau'^A,\tau''^A)\;, 
\end{eqnarray}
where $\alpha=a,a'$ and the arrow indicates time ordering in the
expansion of the exponential. Analogously
\begin{eqnarray}
&&U_\beta(t',t'')\equiv\bar
U_\beta(\tau'^B,\tau''^B)\;, 
\end{eqnarray}
with $\beta=b,b'$. Now Eq. (\ref{evol3}) can be rewritten as
\begin{eqnarray}
|\Psi(t)\rangle&=&\bar U_{b'}(\tau^B,\tau_r^B)\;\bar
U_b(\tau_r^B,0)
\;\times\nonumber\\
&&\bar U_{a'}(\tau^A,\tau_s^A)\; U_c(t_r^B,t_s^A)\;
\bar U_a(\tau_s^A,0)\;|\Psi\rangle
\;\label{evol3bis}.
\end{eqnarray}
Notice that the state $|\Psi(t)\rangle$ in (\ref{evol3bis}) depends on
$t_0^A$, $t_s^A$, $t_0^B$ and $t_r^B$ through PTIs and through the
term $U_c(t_r^B,t_s^A)$, defined in (\ref{udic}).  As already
discussed in the previous section, the random phases $\varphi_e$
present in (\ref{udic}) prevents Bob from recovering the CTI
transit time $t_r^B-t_s^A$.

This example may be easily generalized to the case of multiple
exchanges. Define $t_h^A$ and $t^B_h$ the times at which the last
change in Alice and Bob's Hilbert space took place, {\it i.e.}  the
last time at which they either sent or received a signal. Expressing
it in terms of the PTIs $\tau_h^A=t_h^A-t_0^A$ and
$\tau_h^B=t_h^B-t_0^B$, the state of the system is then
\begin{eqnarray}
|\Psi(t)\rangle&=&\bar U_{A}(\tau^A,\tau_h^A)\;\bar
U_B(\tau^B,\tau_h^B)\; U_{C}(t,t_h)\;|\bar\Psi\rangle
\;\label{generalizz},
\end{eqnarray}
where $A$, $B$ and $C$ refer respectively to the Hilbert spaces of
Alice, Bob and the exchanged system at time $t$, and $t_h$ is the last
time in which the Hilbert space of the exchanged system has been
modified. As can be seen by iterating Eq. (\ref{evol3bis}), the state
vector $|\bar\Psi\rangle$ in Eq. (\ref{generalizz}) depends only on
PTIs and on the transit times of the systems Alice and Bob have
exchanged. To show that the state $|\Psi(t)\rangle$ of
Eq. (\ref{generalizz}) does not contain useful information to
synchronize their clocks, suppose that (say) Bob performs a
measurement at time $t$. The state he has access to is given by
\begin{eqnarray}
\rho_B(t)&=&{\mbox{Tr}}_{AC}\left[|\Psi(t)\rangle\langle\Psi(t)|
\right]\nonumber\\&=&
\bar
U_B(\tau^B,\tau_h^B)\;{\mbox{Tr}}_{AC}\left[|\bar\Psi\rangle\langle\bar\Psi|
\right]
\;\bar U^\dag_B(\tau^B,\tau_h^B)
\;\label{statob},
\end{eqnarray}
where ${\mbox{Tr}}_{AC}$ is the partial trace over ${\cal H}_C$ and
${\cal H}_A$ and where the cyclic invariance of the trace and the
commutativity of operators acting on different Hilbert space has been
used. The state $\rho_B(t)$ does not depend on $\tau^A$. The only
informations relevant to clock synchronization (that connect events on
Alice's side to events on Bob's side) that may be recovered are the
CTI transit times of the exchanged systems. However, in the case of
complete dephasing ($\epsilon=1$), these quantities are irremediably
spoiled by the random phases as discussed previously.

Up to now we have shown that by exchanging physical systems and
performing a measurement, Alice and Bob cannot recover sufficient
information to synchronize their clocks if the environment is
completely dephasing.  In other words, Alice can always encode some
information on the system she sends Bob, but any operation she did
will always be referred to her PTI and will thus be useless to Bob if
he ignores any CTI. That is equivalent to say that Alice may always
send Bob some photographs of her clock, but Bob will have no use of
them, since he cannot arrange them relative to his own time axis.  A
better strategy could be to measure only part of their systems and
employing post-selection schemes. As will be shown in the next
section, even in this case all their efforts are in vain if
Hyp.~{\ref{dephasing}} applies.

\subsection{Post-Selection schemes}\label{ps}
Allow Alice and Bob to make partial measurements on their systems. The
global system evolution is no longer unitary, since the measurements
will project part of the Hilbert space into the eigenstates of the
measured observable. The communication of the measurement results
permits the implementation of post-selection schemes. We will show
that also in this case, Alice and Bob cannot synchronize their clocks
in presence of dephasing in the communication channel.

Using Naimark extension {\cite{naimark}}, one can assume the
projective-type measurement as the most general. Suppose that Alice
performs the first measurement at time $t_m^A$ on a part of her
system. Define ${\cal H}_{A_1}$ the Hilbert space that describes such
system, so that ${\cal H}_{A}={\cal H}_{A_0}\otimes {\cal H}_{A_1}$ is
the Hilbert space of Alice. The state of the system after the
measurement for $t>t_m^A$ (and before any other measurement or system
exchange) is
\begin{eqnarray}
|\Psi(t)\rangle&=&U(t,t_m^A)\; P(A_{1})\; |{\Psi(t_m^A)}\rangle
\; \label{evolf},
\end{eqnarray}
where $|\Psi(t_m^A)\rangle$ is given in Eq. (\ref{generalizz}) and the
global evolution operator is \begin{eqnarray} U(t,t_m^A)=\bar
U_A(\tau^A,\tau_m^A)\;\bar
U_B(\tau^B,t_m^A-t_0^B)\;U_C(t,t_m^A)\;\label{globalu}
\end{eqnarray}
with $\tau^A_m=t_m^A-t_0^A$.  In Eq. (\ref{evolf}) the measurement
performed by Alice on $|\Psi(t_1^A)\rangle$ is described by the
projection operator
\begin{equation}
P(A_{1}) |\psi\rangle
\equiv \frac{1}{|| \langle \psi|\phi
\rangle_{A_{1}}||} \Big(|\phi\rangle_{A_{1}}\langle \phi | \otimes
\openone_{{A}_{0}}\Big)|\psi\rangle
\;\label{proiettore},
\end{equation} 
where $\openone_{{A_0}}$ is the identity on ${\cal H}_{A_0}$,
$|\phi\rangle_{A_{1}}\in {\cal H}_{A_{1}}$ is the eigenstate relative
to Alice's measurement result $\phi$. Notice that
Eqs. (\ref{evolf})--(\ref{proiettore}) take into account the
post-selection scheme in which Alice communicates her measurement
result to Bob, since the operator $U(t,t_m^A)$ can depend on Alice's
measurement result $\phi$. Using again the commutation properties
between operators that act on different spaces, Eq. (\ref{evolf})
simplifies to \begin{eqnarray} |\Psi(t)\rangle&=&\bar
U_B(\tau^B,\tau_h^B)\;U_C(t,t_h)\;\bar
U_A(\tau^A,\tau_m^A)\nonumber\\&&
\; P(A_{1})\;\bar U_A(\tau_m^A,\tau_h^A) |\bar{\Psi}\rangle
\;\label{finito}.
\end{eqnarray}
Eq. (\ref{finito}) shows that even though the partial measurement
introduces a non-unitary evolution term, this allows Alice to encode
in the state only information about her PTI $\tau_m^A$ and nothing on
the absolute time $t_m^A$ (as measured by an external clock) or on any
CTI. In fact, the same considerations of Eq. (\ref{statob}) apply and
no information relevant to clock synchronization can be extracted from
the state (\ref{finito}). The formalism introduced allows also to
consider the situation in which Alice does not look at her results (or
does not communicate them to Bob): in this case, in Eq. (\ref{finito})
one must perform the sum on all the possible measurement results
weighted by their outcome probability.

In the most general scenario Alice and Bob will perform multiple
partial measurements, communicate by exchanging physical systems (as
analyzed in the previous section) and again perform partial
measurements. By iterating (\ref{finito}) one can show that none of
these efforts allows them to extract any CTI.

Before concluding, it is worth to comment how the quantum clock
synchronization scheme proposed in {\cite{jozsa}} is related with our
analysis. In {\cite{jozsa}}, the authors assume as a starting point
that Alice and Bob share an entangled state of the form
\begin{eqnarray} |\chi\rangle=\sum_{a,b}\chi_{ab}|a\rangle\>|b\rangle
\;\label{statojosza},
\end{eqnarray}
where $|a\rangle$ and $|b\rangle$ are energy eigenstates of Alice and
Bob's systems respectively, and where the sum on the indexes $a$ and
$b$ runs over non-degenerate eigenstates.  From the considerations
given in the present section, one can show that, in the presence of a
dephasing channel, such a state cannot be obtained starting from the
initial state given in Eq. (\ref{statoiniz}) without introducing in it
some stochastic phases. For this reason, it cannot be obtained without
relaxing Hyp.~{\ref{dephasing}}: such a protocol is then equivalent to
classical protocols {\cite{comment}}. In fact, if one relaxes
the hypotheses of channel dephasing, then it is possible to achieve
also classical clock synchronization.

\section{Conclusion}\label{concl}
In conclusion, a definition of clock synchronization was given and it
was shown that, under some very general hypotheses that preclude the
possibility of employing classical protocols, such a synchronization
is not possible. This does not imply that quantum mechanics may not be
exploited in the clock synchronization procedures, but it may be
limited only to enhancing classical clock synchronization protocols
{\cite{chuang,altro,prl}}. Indeed we have shown elsewhere {\cite{prl}}
that quantum mechanics may be used to cancel the effect of dispersion
in clock synchronization.

\acknowledgments
We wish to acknowledge support from DARPA grant {\#} F30602-01-2-0546
under the QUIST program, ARO grant {\#} DAAD19-001-0177 under the MURI
program, and NRO grant {\#} NRO-000-00-C-0158.
\vskip 1\baselineskip
\par\noindent
{\footnotesize
$^*$ Email address: vittorio@mit.edu \par\noindent
$^\dag$ Email address: slloyd@mit.edu \par\noindent
$^\ddagger$ Email address: maccone@mit.edu \par\noindent
$^\S$ Email address: smshahri@mit.edu
}

\narrowtext


\begin{references}
\bibitem{nota} Consider, for example, a time dependent perturbation
$H'_C(t)$ that acts on the system for a time $\delta t$ during the
interval $[t_s,t_r]$, taking the value $\hbar\sum_e P_e\nu_e$. Since
$H'_C(t)$ is diagonal in the energy basis of $H^o_C$, then it is easy
to derive the evolution dynamics (\ref{udic}).
\end{references}
\end{document}